\documentclass[12pt,preprint,notitlepage,superscriptaddress]{revtex4-2}

\usepackage[utf8]{inputenc}
\usepackage{textgreek}
\DeclareUnicodeCharacter{03BC}{\textmu}
\usepackage{amsmath, orcidlink}
\usepackage{booktabs}
\usepackage{graphicx}
\usepackage{xcolor}
\usepackage{url}
\usepackage{xr}
\usepackage{etoolbox}

\makeatletter
\patchcmd{\frontmatter@abstract@produce}
  {\penalty-200\relax}
  {\penalty10000\relax}
  {}
  {}
\makeatother

\begin{document}

\title{How significant is the lensing interpretation of GW231123?}

\author{Xikai Shan \orcidlink{0000-0003-3201-061X}}
% \email{xikai_shan@mail.bnu.edu.cn} 
\affiliation{Department of Astronomy, Tsinghua University, Beijing 100084, China}
\affiliation{Max Planck Institute for Gravitational Physics (Albert Einstein Institute) \\
Am Mühlenberg 1, D-14476 Potsdam-Golm, Germany}

\author{Huan Yang \orcidlink{0000-0002-9965-3030}}
\email{hyangdoa@tsinghua.edu.cn} \affiliation{Department of Astronomy, Tsinghua University, Beijing 100084, China}

\author{Shude Mao \orcidlink{0000-0001-8317-2788}}
\affiliation{Department of Astronomy, Westlake University, Hangzhou 310030, Zhejiang Province, China}

\begin{abstract}
GW231123 is one of the most unusual gravitational-wave (GW) events, with exceptionally large inferred masses and near-extremal spins, offering an opportunity to test whether propagation effects contribute to these properties.
We therefore examine whether the data support wave-optics microlensing embedded in a strong-lensing galaxy, whose detection becomes increasingly likely as observations accumulate, whether this interpretation can explain these properties, and how significant the preference remains under detector noise and waveform systematics.
We compare six hypotheses: unlensed, isolated point mass, and embedded point-mass (EPM) and binary-lens (EB) effective models in Type-I (minimum) and Type-II (saddle) macro images.
The EB Type-I model is most favored.
For the most accurate waveform model NRSur7dq4, it gives $\log_{10}B^{\rm EB-I}_{\rm U}=2.60$, versus $0.89$ for Type II, indicating sensitivity to macro-image geometry.
Within Type I, however, the binary improves over the point mass by only $\log_{10}B^{\rm EB-I}_{\rm EPM-I}=0.16$ and $\Delta\ln\mathcal{L}_{\max}=0.56$, providing no clear evidence for structure beyond a single effective perturber.
Moreover, under embedded lensing, waveform-template discrepancies and inferred masses and spins are reduced. However, real O4a backgrounds from numerical-relativity injections show that the apparent lensing evidence is sensitive to waveform systematics and realistic detector noise: although the commonly used waveform IMRPhenomXPHM gives the largest Bayes factor, $\log_{10}B^{\rm EB-I}_{\rm U}=4.52$, it is less exceptional relative to its own background, with a false-alarm probability of $6.5$--$8\%$, whereas NRSur7dq4 gives only $2$--$3\%$. Thus, waveform systematics can amplify apparent lensing evidence, but GW231123 remains an intriguing lensing candidate.
\end{abstract}

\maketitle

On November 23, 2023, the LIGO-Virgo-KAGRA (LVK)~\citep{Aasi_2015, Acernese_2015, KAGRA:2020tym} collaboration detected the gravitational-wave (GW) signal GW231123\_135430 (hereafter GW231123), which originated from the merger of a binary black hole (BBH) system.
The black holes had masses of $137^{+22}_{-17}M_\mathrm{\odot}$ and $103^{+20}_{-52}M_\mathrm{\odot}$ (at 90\% credible intervals) and very high spins of $0.90^{+0.10}_{-0.19}$ and $0.80^{+0.20}_{-0.51}$. 
These features are unusual, particularly because the secondary mass falls within the theoretical pair-instability mass gap (about 60--130 $M_\mathrm{\odot}$~\citep{2019ApJ...887...53F, 2020ApJ...902L..36F, 2021ApJ...912L..31W, 2023MNRAS.526.4130H}) and the spins are near the extremal Kerr limit~\citep{Kerr:1963ud, PhysRev.174.1559}.
Therefore, it is important to verify the true nature of this signal, because confirming these parameters would challenge current stellar evolution theory or require non-standard formation channels.

Numerous scenarios have been proposed to explain GW231123, including hierarchical mergers~\citep{Kiroglu:2025vqy, Li:2025pyo, Paiella:2025qld, Passenger:2025acb, Mai:2025jmk, Liu:2025yok, Li:2025fnf}, in which the high-spin subpopulation is often associated with the hierarchical merger channel~\citep{Li:2023yyt}; active galactic nucleus (AGN) disk-assisted mergers~\citep{Bartos:2025pkv,Delfavero:2025lup}; primordial black holes (PBHs)~\citep{Yuan:2025avq, DeLuca:2025fln}; cosmic string collapse~\citep{Cuceu:2025fzi}; collisions of exotic compact objects such as boson stars~\citep{LIGOScientific:2025rsn}; overlapping signals~\citep{Hu2025}; instrumental issues such as microglitches~\citep{ray2025gw231123extremespinsmicroglitches}; the presence of multiple quasinormal modes~\citep{siegel2025gw231123ringdowninterpretationmultimodal}; and gravitational lensing~\citep{Hannuksela2019SearchEvents, Abbott2021SearchRun, Abbott2024SearchNetwork, o4alensing, Wright2022Gravelamps:Selection, Li:2022izh, Li:2022grj, Guo:2022dre, Liu:2023ikc, Chakraborty_2025_a, Chakraborty_2025_b, Seo_2025, Goyal2025, Liu2025GW231123, chan2025discoveringgravitationalwaveformdistortions}.
% \hy{This paragraph and the following one should be used to build up the tension: why previous literature is not good/enough?} \hy{The last sentence should be removed as it just says what we do but not why we do it.}

However, these hypotheses also face distinct challenges in explaining GW231123. 
% For the hierarchical-merger interpretation, \citet{Passenger:2025acb} found that GW231123 can be produced within a specific $(2\mathrm{G}+2\mathrm{G})$ merger model in a dense star cluster at the $\sim 5\%$ level. If one further folds this number with the recent GWTC-4-based estimate that higher-generation BBH mergers constitute only $2.5^{+5.5}_{-1.8}\%$ of the overall BBH population~\citep{ray2026astrophysicaloriginbinaryblack}, and with theoretical expectations that $(2\mathrm{G}+2\mathrm{G})$ mergers are typically suppressed by at least an additional factor of 10 relative to the dominant 2G+1G channel~\citep{vijaykumar2026maximummassratiohierarchical}, then this implies that the fraction of such GW231123-like events in the total BBH population is only of order $10^{-4}$ or less.
For the hierarchical-merger interpretation, \citet{Passenger:2025acb} found that the unusually large component spins of GW231123 are possible but relatively uncommon within a specific $(2\mathrm{G}+2\mathrm{G})$ model in dense star clusters, with $p\simeq5\%$. Importantly, this test considers the component spins and does not require the model to reproduce the component masses. Moreover, the required $(2\mathrm{G}+2\mathrm{G})$ configuration is itself expected to be rare. Dense-cluster models predict 2G+1G to be the dominant hierarchical-merger channel, while mergers involving higher-generation black holes are expected to occur at substantially lower rates~\citep{vijaykumar2026maximummassratiohierarchical}. Even in globular-cluster simulations with zero black-hole birth spins, which favor remnant retention and repeated mergers, only about $1\%$ of all cluster mergers and $3\%$ of detectable cluster mergers are $(2\mathrm{G}+2\mathrm{G})$ systems~\citep{Rodriguez:2019huv}. Nonzero birth spins further suppress repeated mergers through gravitational recoil~\citep{Rodriguez:2019huv}. Consistently, under a recent Gravitational-Wave Transient Catalog 4 (GWTC-4)-based three-channel interpretation, the entire higher-generation subpopulation contributes only $2.5^{+5.5}_{-1.8}\%$ of local BBH mergers~\citep{ray2026astrophysicaloriginbinaryblack}. Thus, although hierarchical assembly remains possible, explaining GW231123 in this way requires a very rare $(2\mathrm{G}+2\mathrm{G})$ configuration.

For AGN-based interpretations, GW231123 requires either sustained coherent gas accretion or very high-generation mergers such as $(4\mathrm{G}+3\mathrm{G})$~\citep{Bartos:2025pkv,Delfavero:2025lup}. Moreover, the predicted rates depend strongly on the AGN lifetime, disk model, and black-hole initial mass function, making this interpretation sensitive to specific AGN conditions and model assumptions.
For PBH scenarios, reproducing GW231123 requires a relatively large PBH abundance close to current exclusion bounds from the cosmic microwave background and X-ray observations, while the extreme spins may require additional accretion, leaving the interpretation in tension with existing constraints~\citep{Yuan:2025avq}.
Cosmic-string collapse is strongly disfavored, as Bayesian model comparisons favor a BBH origin over a cosmic-string signal~\citep{Cuceu:2025fzi}.
Collisions of exotic compact objects, such as boson stars, can mimic the short, ringdown-dominated waveform of a high-mass BBH merger. However, the LVK analysis finds the high-mass BBH interpretation to have a higher astrophysical probability because GW231123 is consistent with the rates and properties of the currently understood BBH population~\citep{LIGOScientific:2025csr,LIGOScientific:2025rsn}.
% Finally, explaining the signal with instrumental glitches (microglitches) would require near-simultaneous unmodeled noise in multiple detectors, which is statistically dubious~\citep{ray2025gw231123extremespinsmicroglitches}.

Gravitational lensing provides a viable alternative for GW231123. Unlike the source-formation scenarios discussed above, lensing does not require modifying the intrinsic formation history of the binary, but instead acts as a propagation effect with a non-negligible lensing probability. The most recent lensing analysis of the first part of the fourth observing run (O4a) estimates a typical relative strong-lensing rate of order $10^{-3}$ for detectable GW events~\citep{o4alensing}, while theoretical studies likewise suggest that the probability of observing the first lensed GW increases rapidly as GW observations accumulate~\citep{Li:2018prc, Oguri:2018muv, Yang:2021viz, Xu:2021bfn}. More importantly, lensing offers an event-level test that is complementary to formation- and population-based arguments: in the wave-optics regime, microlensing produces characteristic frequency-dependent amplitude and phase modulations that can be searched for directly in an individual GW waveform~\citep{Takahashi:2003ix, Shan:2023qvd, Shan:2025jpt}. Indeed, GW231123 stands out in the O4a single-event search as the only clear outlier showing support for lensing-induced waveform distortions~\citep{o4alensing}. If lensing is not modeled, these distortions can also bias the inferred masses and spins. Lensing therefore provides an astrophysically plausible and directly testable explanation for the unusual inferred properties of GW231123, without requiring exotic source physics or new fundamental physics.

Recent lensing analyses of GW231123 have found support for wave-optics distortions~\citep{o4alensing,Goyal2025,Liu2025GW231123}. In addition, phenomenological millilensing analyses suggest a three-image configuration (private communication~\citep{Liu2025GW231123}), which cannot be produced by a standard isolated point-mass lens. This indicates that, if GW231123 is lensed, the lens system may be more complex than a single isolated object. 
Since microlenses generally reside in foreground lensing galaxies, a realistic microlensing model should include the external strong-lensing field. In such environments, the stellar field may contain many microlenses and generate a complex network of microimages, making direct parameter estimation of the full stellar population computationally prohibitive. We therefore use the embedded point-mass (EPM) and embedded binary-lens (EB) models as low-dimensional effective descriptions of the local stellar field.
The EPM model captures the dominant effective wave-optics response through a single perturber embedded in the macro-lensing field, while the EB model introduces a second effective perturber to allow for additional local structure. This interpretation is motivated by \citet{Goyal2025}, who showed that the wave-optics response of a dense field of lower-mass microlenses can be closely reproduced by a single heavier microlens. They also found that embedding a point microlens in an external strong-lensing potential improves the maximum-likelihood fit to GW231123 relative to an isolated point mass. We therefore adopt the EPM model as the simplest effective description of embedded microlensing and use the EB model to test whether the data require additional structure beyond a single effective perturber.
We stress that these models are effective descriptions rather than literal one- and two-object configurations, as such configurations are strongly disfavored by astrophysical probability considerations~\citep{Cheung:2026pky,vujeva2026probabilitygravitationalwavelensingintermediatemass}.

However, even with these low-dimensional effective models, computational cost remains a major obstacle. Unlike an isolated point-mass lens, embedded microlensing models generally have no analytic wave-optics solution and require expensive numerical diffraction calculations, making direct Bayesian parameter estimation impractical. To overcome this difficulty, we construct fast neural-network surrogates for embedded microlensing in both Type-I (minimum) and Type-II (saddle) macro-image geometries, enabling the wave-optics amplification to be evaluated within milliseconds. 
The neural-surrogate architecture and its test-set accuracy are shown in Extended Data Figures~\ref{fig:transformer} and~\ref{fig:app2}, respectively.
The accuracy of the EPM approximation is further validated in Extended Data Figure~\ref{fig:epm_mismatch_histogram_typei_typeii}.
With these surrogates, we systematically compare six hypotheses: unlensed (U), isolated point mass (PM), and the EPM and EB effective models in Type-I and Type-II macro images.

Because an apparent preference for lensing may also arise from detector noise or waveform-model systematics, we further assess its significance using event-matched backgrounds. We inject unlensed direct numerical-relativity signals into real O4a detector noise and analyse them with the same parameter-estimation framework, allowing the observed lensing preference to be calibrated against realistic noise fluctuations and waveform systematics.

% We find that the embedded Type-I models receive the strongest support. For NRSur7dq4, the embedded binary Type-I model gives $\log_{10}B^{\rm EB-I}_{\rm U}=2.60$, compared with $0.89$ for Type II, indicating that GW231123 contains information sensitive to the macro-image geometry. Within Type I, however, the embedded binary improves over the embedded point-mass model by only $\log_{10}B^{\rm EB-I}_{\rm EPM-I}=0.16$, \codex{providing no clear evidence for additional structure beyond a single dominant perturber.}
% Moreover, under embedded lensing, the BBH component-mass posteriors become more consistent, with lower masses and spin-magnitude medians. However, the real-O4a backgrounds show that a larger lensing Bayes factor need not imply a more exceptional event: IMRPhenomXPHM gives $\log_{10}B^{\rm EB-I}_{\rm U}=4.52$, yet with a false-alarm probability of \codex{$6.5$--$8\%$}, whereas NRSur7dq4 gives only $2$--$3\%$. Thus, waveform systematics can amplify apparent lensing evidence, while GW231123 remains an intriguing lensing candidate.
We find that the Type-I embedded models receive the strongest support. For NRSur7dq4, EB-I gives $\log_{10}B^{\rm EB-I}_{\rm U}=2.60$, compared with $0.89$ for EB-II, indicating sensitivity to the macro-image geometry. Within Type I, however, the evidence gain of EB-I over EPM-I is small, providing no clear evidence for additional structure beyond a single effective perturber. Embedded lensing also shifts the inferred source properties towards lower masses and less extreme spins. Finally, event-matched backgrounds show that the larger lensing Bayes factor obtained with IMRPhenomXPHM does not necessarily correspond to a more exceptional event relative to its own background, highlighting the importance of waveform systematics and realistic detector noise in assessing the lensing significance of GW231123.

\section*{Results}

To assess the impact of microlensing under different lensing configurations, we incorporate the isolated and embedded microlensing templates into the \texttt{Bilby} inference pipeline~\citep{bilby_paper,2020MNRAS.499.3295R}. We consider six hypotheses: an unlensed model (U), an isolated point-mass lens (PM), an embedded point-mass lens in Type I and Type II macro images (EPM-I and EPM-II), and an embedded binary lens in Type I and Type II macro images (EB-I and EB-II). To examine the robustness of the results against waveform systematics, we repeat the inference using three waveform models: \texttt{NRSur7dq4} (NRSur), \texttt{IMRPhenomXPHM} (Phenom), and \texttt{SEOBNRv5PHM} (EOB).

In this work, we use publicly available time-series data for LIGO Hanford (H1) after glitch subtraction and LIGO Livingston (L1) from the Gravitational Wave Open Science Center\footnote{\url{www.gw-openscience.org}}. The noise power spectral density (PSD), detector noise calibration envelopes, and other metadata are downloaded from Zenodo\footnote{\url{https://zenodo.org/doi/10.5281/zenodo.15832843}}. We use the same configuration, including the GW prior ranges and data segments, as described in~\cite{LIGOScientific:2025rsn}, and use the \texttt{dynesty} sampler~\citep{Speagle_2020}.

For the model comparison, we consider the unlensed, isolated point-mass, and embedded Type-I and Type-II point-mass and binary hypotheses introduced above. The amplification factor for the isolated point-mass model is computed with \texttt{GLoW}~\citep{Villarrubia_Rojo_2025}, while those for the embedded models are evaluated using our Type-I and Type-II neural-network surrogates. Details of the lens models, surrogate construction, priors, and Bayesian inference are provided in Methods.

Table~\ref{tab:sum} reports the lensing Bayes factor and maximum-likelihood changes for the five lensing hypotheses relative to the unlensed hypothesis under the same waveform template.

\begin{table*}[h!]
    \centering
    \renewcommand{\arraystretch}{1.15}
    \caption{Model comparison for the five lensing hypotheses. Each entry gives $\log_{10}B^{H}_{\rm U}$, followed in parentheses by $\Delta\ln\mathcal{L}_{\max}$ relative to the unlensed result obtained with the same waveform template.}
    \begin{tabular}{lccc}
    \toprule
    Hypothesis & NRSur & Phenom & EOB \\
    \midrule
    PM     & $1.47\ (3.19)$  & $3.54\ (8.37)$  & $-0.05\ (2.28)$ \\
    EPM-I  & $2.44\ (6.32)$  & $4.47\ (12.41)$ & $0.61\ (5.20)$ \\
    EB-I   & $2.60\ (6.87)$  & $4.52\ (12.14)$ & $0.73\ (6.11)$ \\
    EPM-II & $1.45\ (1.70)$  & $3.33\ (7.04)$  & $0.55\ (4.23)$ \\
    EB-II  & $0.89\ (3.47)$  & $3.30\ (9.59)$  & $0.39\ (5.52)$ \\
    \bottomrule
    \end{tabular}
    \label{tab:sum}
\end{table*}

Based on these results, we draw two main conclusions. First, the EB-I hypothesis has the highest evidence across all three waveform analyses. Focusing on NRSur7dq4 as our primary high-fidelity waveform model, we find $\log_{10}B^{\rm EB-I}_{\rm U}=2.60$, compared with $0.89$ for EB-II, favoring a Type-I macro-image geometry. This suggests that GW231123 contains signatures sensitive to the macro-image geometry. Second, within Type I, EB-I further improves over EPM-I by $\log_{10}B^{\rm EB-I}_{\rm EPM-I}=0.16$ and $\Delta\ln\mathcal{L}_{\max}=0.56$. However, this improvement is limited, and the evidence does not clearly favor additional local structure beyond the EPM description.

\begin{figure}[h!]
    \centering
    \includegraphics[width=\linewidth]{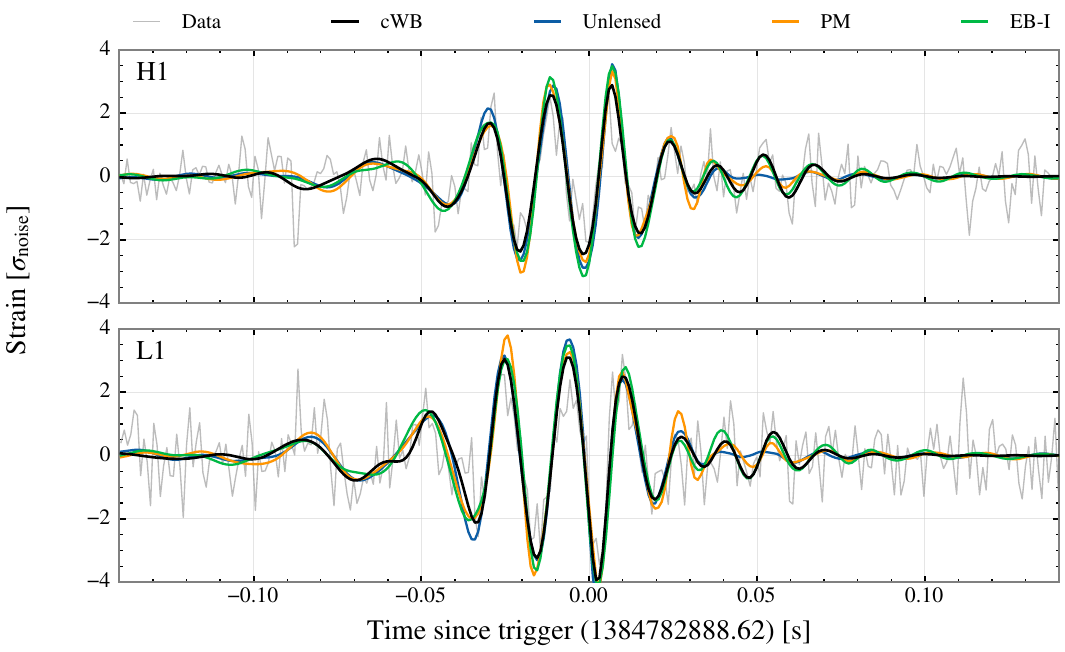}
    \caption{Time-domain whitened strain in H1 and L1, in units of the noise standard deviation. The curves show the observed data, the \texttt{cWB} reconstruction, and the NRSur maximum-likelihood waveforms under U, PM, and EB-I.}
    \label{fig:fig1}
\end{figure}

Figure~\ref{fig:fig1} compares the observed whitened strain and coherent WaveBurst (\texttt{cWB}) reconstruction with the NRSur maximum-likelihood waveforms for U, PM, and EB-I. The data and \texttt{cWB} reconstruction are those reported in~\citet{LIGOScientific:2025rsn} and released on Zenodo~\footnote{\url{https://zenodo.org/doi/10.5281/zenodo.15832843}}.
The waveforms cover 20--448 Hz, matching the inference band. As a template-free method, \texttt{cWB} identifies coherent power across the detector network~\citep{2008CQGra..25k4029K}. The EB-I waveform tracks portions of the reconstructed morphology that are not captured by the isolated point-mass waveform.

\begin{figure}[h!]
    \centering
    \includegraphics[width=\linewidth]{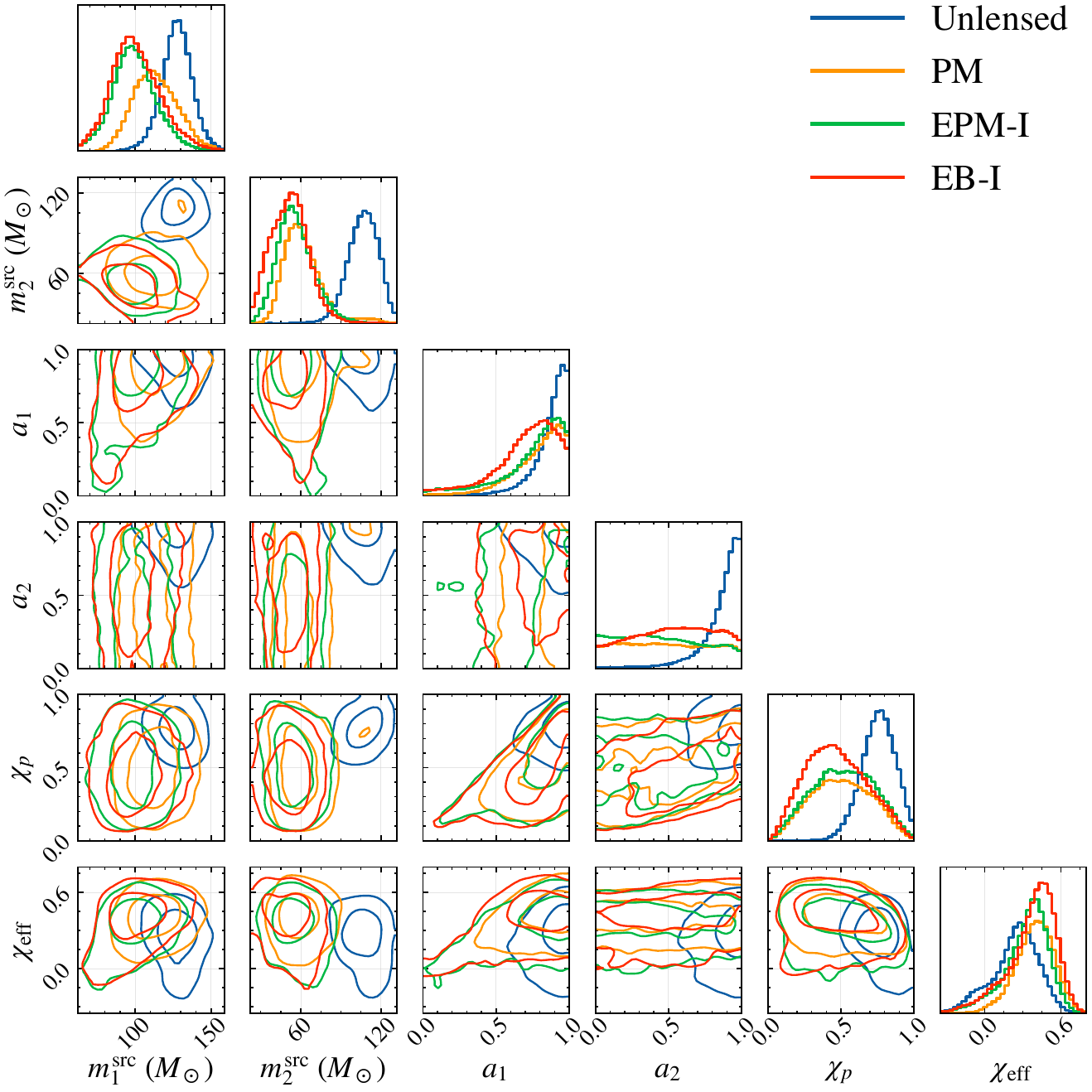}
    \caption{NRSur posterior distributions for the source-frame component masses, spin magnitudes, and effective spin parameters under U, PM, EPM-I, and EB-I. Contours enclose 50\% and 90\% credible regions.}
    \label{fig:fig2}
\end{figure}

Figure~\ref{fig:fig2} and Extended Data Table~\ref{tab:source_params_comparison} compare the NRSur source posteriors under U, PM, EPM-I, and EB-I. One can see that lensing shifts the inferred source-frame masses downward: the median component masses change from $(127.7,107.1)\mathrm{M}_\odot$ for U to $(112.7,59.5)\mathrm{M}_\odot$ for PM and $(99.2,51.8)\mathrm{M}_\odot$ for EB-I. The median spin magnitudes also decrease, making the inferred spins less extreme.

However, Extended Data Figure~\ref{fig:source_mass_appendix}, which shows the source-frame component-mass posteriors for U, PM, EPM-I, EB-I, EPM-II, and EB-II using the NRSur, Phenom, and EOB waveform models, reveals an interesting behavior across different waveform models. Under the unlensed hypothesis, NRSur and EOB are in good agreement, whereas Phenom shows a noticeable offset. When an isolated point-mass lens is introduced, this inconsistency among the waveform models is not removed; instead, EOB becomes the most discrepant of the three. In contrast, when embedded lensing is considered, the discrepancies among the waveform models are reduced to varying degrees.
This may indicate that part of the waveform-model inconsistency under the unlensed hypothesis is associated with unmodeled lensing signatures.

\begin{figure}[h!]
    \centering
    \includegraphics[width=\linewidth]{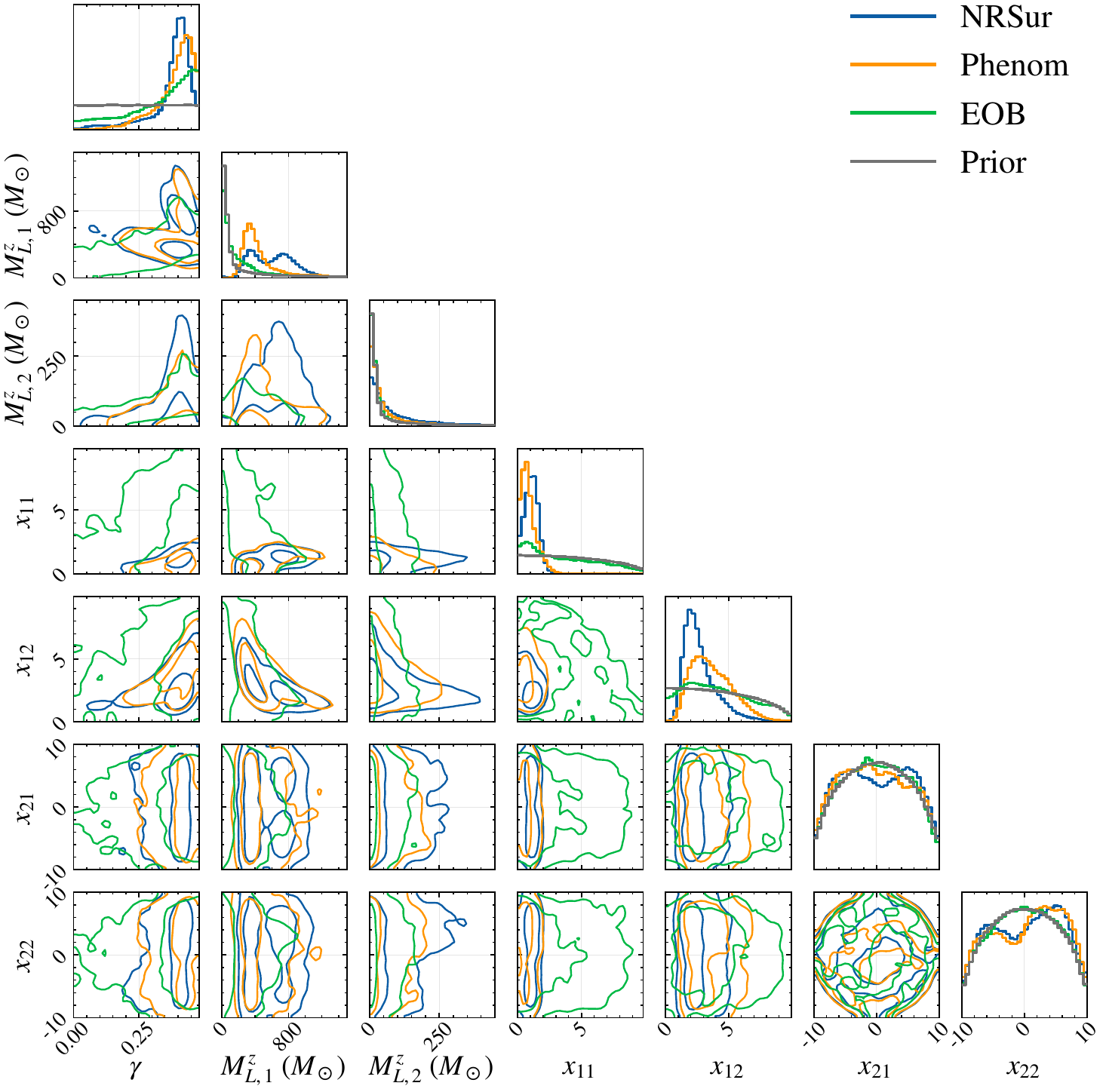}
    \caption{EB-I posterior distributions for the external shear, redshifted masses, and lens-plane coordinates of the two effective perturbers obtained with NRSur, Phenom, and EOB. Gray curves in the one-dimensional panels show the sampling priors; contours enclose 50\% and 90\% credible regions.}
    \label{fig:fig3posterior}
\end{figure}

Figure~\ref{fig:fig3posterior} compares the EB-I lens posteriors obtained with NRSur, Phenom, and EOB with the corresponding sampling priors. The NRSur and Phenom results are broadly consistent with each other and give narrower posteriors than EOB for several lens parameters, while the parameters associated with the secondary effective perturber remain weakly constrained.

Taking NRSur as our reference waveform model, we infer $\gamma=0.40^{+0.06}_{-0.22}$ and, adopting the singular isothermal ellipsoid model $\kappa=\gamma$~\citep{1994A&A...284..285K}, $\mu=4.97^{+6.09}_{-3.43}$. The corresponding redshifted effective perturber masses are $622^{+455}_{-357}\mathrm{M}_\odot$ and $52^{+314}_{-47}\mathrm{M}_\odot$. 
One can find that the parameters associated with the secondary effective perturber show only minor departures from their sampling priors. Together with the small evidence gain of EB-I over EPM-I, this provides no clear evidence for additional structure beyond a single effective perturber. 
The numerical summaries are given in Extended Data Table~\ref{tab:lens_params}.

However, a Bayes factor only quantifies the relative support between specified hypotheses under the assumed waveform models, priors, and likelihood. 
In GW lensing analyses, such a preference can also be enhanced by waveform-model systematics and fluctuations in realistic detector noise. We therefore quantify the false-alarm probability (FAP), defined here as the fraction of unlensed events that produce lensing evidence at least as large as that observed for GW231123.

To estimate this event-conditioned FAP, we construct a null background using 100 injections into selected real O4a H1/L1 noise realizations. The injected signal is taken directly from a numerical-relativity simulation of an unlensed GW231123-like source, while the sky geometry and network signal-to-noise ratio (SNR) are matched to the adopted GW231123 reference. Each injection is recovered under the U and EB-I hypotheses, with EB-I being the most favored lensing model, using the same production NRSur parameter-estimation pipeline as for GW231123. These backgrounds therefore capture both differences between the direct numerical-relativity injections and the recovery waveform model, representing waveform systematics, and the effects of realistic non-Gaussian and non-stationary detector noise.
A representative background injection and its corresponding NRSur and Phenom waveform comparisons are shown in Extended Data Figure~\ref{fig:appendix_figure3_background_injection_nrsur_best_fit}.

In addition, we perform 18 corresponding trials with Phenom recoveries. To limit the additional computational resources devoted to this auxiliary cross-check, we restrict the Phenom background analysis to the 18 completed trials, which is sufficient to reveal a clear qualitative trend: all 18 trials yield larger background Bayes factors than their corresponding NRSur recoveries.

\begin{figure}[h!]
    \centering
    \includegraphics[width=\linewidth]{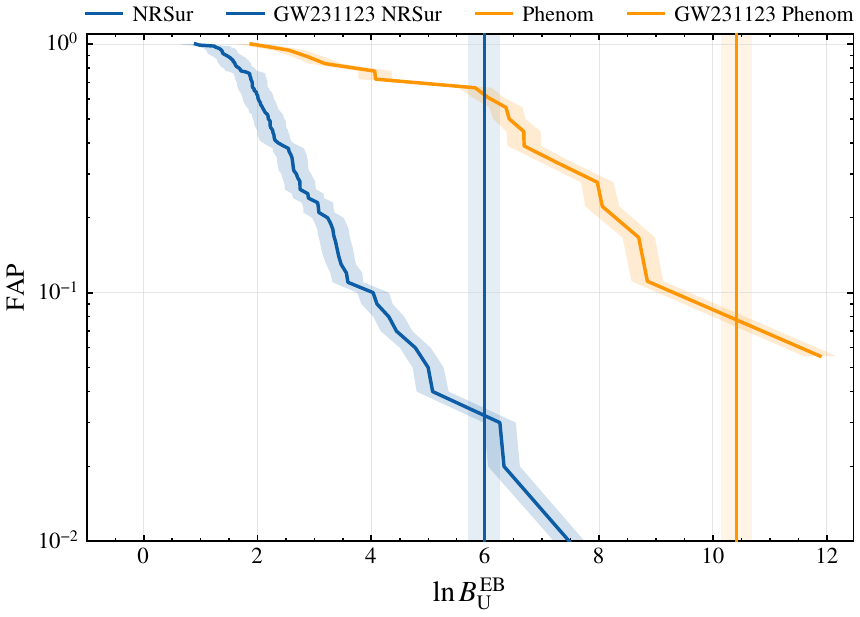}
    \caption{EB-I versus U background distributions from 100 NRSur (blue) and 18 Phenom (orange) real-O4a trials. Solid curves show the complementary cumulative distribution functions (CCDFs), while the shaded regions indicate the 68\% uncertainty from Bayes-factor errors. Vertical lines and bands mark the GW231123 measurements and their 68\% uncertainties.
}
    \label{fig:background}
\end{figure}

Figure~\ref{fig:background} shows the background distributions of $\log_{10}B^{\rm EB-I}_{\rm U}$ from the 100 completed NRSur trials, together with the 18 completed Phenom trials. Taking into account the uncertainty in the Bayes-factor estimate for each event, the FAP for the NRSur analysis is $2$--$3\%$ within the 68\% interval. For Phenom, the corresponding estimate is higher, at $6.5$--$8\%$. Thus, despite its larger observed Bayes factor, the Phenom result is less exceptional relative to its own background than the NRSur result.

This comparison highlights the impact of waveform systematics on the inferred lensing evidence and the importance of including them, together with realistic detector noise, when assessing the FAP for GW231123. Nevertheless, after taking into account waveform-model systematics and realistic non-Gaussian and non-stationary detector noise, the NRSur analysis still yields an FAP of only $2$--$3\%$, suggesting that GW231123 remains an intriguing lensing candidate.

\section*{Discussion}

In this work, we revisited GW231123, which was originally reported with unusual properties, including exceptionally large component masses and near-extremal spins. The important question is whether these properties are intrinsic to the source, and therefore require an unusual formation history, or whether they can be partly produced by propagation effects. Previous studies have shown that an isolated point-mass lens can improve the fit to GW231123 and make its inferred properties somewhat less extreme, but such a model represents only the simplest microlensing configuration and becomes implausible once the population probability is taken into account~\citep{Cheung:2026pky}. 
In addition, microlenses generally reside in foreground lensing galaxies, where an external strong-lensing field is naturally present~\citep{o4alensing,Li:2018prc,Oguri:2018muv,Yang:2021viz,Xu:2021bfn}. It is therefore important to take the external strong-lensing field into account. We thus considered the EPM and EB models in both Type-I and Type-II macro-images, and compared them with the unlensed and isolated point-mass hypotheses.
It is worth noting that we do not claim that the strong-lensing galaxy contains only a single point-mass lens or a binary lens. Instead, a realistic stellar microlensing field may contain many microlenses and generate a large number of microimages~\citep{1986ApJ...306....2K}. The EPM and EB models used here should therefore be regarded as low-dimensional effective descriptions of such a field, with EPM capturing a dominant wave-optics response and EB allowing for additional local structure.
Recent population calculations show that lensing by a literal intermediate-mass black hole hosted by a globular cluster is astrophysically rare and is disfavored as an interpretation of GW231123~\citep{vujeva2026probabilitygravitationalwavelensingintermediatemass}. 
Our inferred effective-perturber masses should not be interpreted as the masses of individual compact objects, so this population prior does not directly apply to the EPM and EB descriptions considered here.

Our analysis shows that the lensing preference depends strongly on the lens environment. For NRSur, the embedded Type-I models are favored over both the isolated point-mass model and their Type-II counterparts, with EB-I giving the largest evidence, $\log_{10}B^{\rm EB-I}_{\rm U}=2.60$. In comparison, the corresponding Type-II result is only $\log_{10}B^{\rm EB-II}_{\rm U}=0.89$. A similar preference for embedded Type-I lensing is also found with the Phenom and EOB waveform models. Therefore, the qualitative ordering that EB-I is the most favored model and that Type-I macro-image geometries are preferred over their Type-II counterparts is consistent across the waveform models considered here.

Regarding the detailed lens structure, the difference between EPM-I and EB-I is much weaker. EPM captures the dominant wave-optics response through a single effective perturber, whereas EB allows for additional local structure. For NRSur, the improvement from EPM-I to EB-I is only $\log_{10}B^{\rm EB-I}_{\rm EPM-I}=0.16$, with $\Delta\ln\mathcal{L}_{\max}=0.56$. Moreover, the parameters associated with the secondary effective perturber show only minor departures from their sampling priors. We therefore find no clear evidence for additional structure beyond a single effective perturber, given the current SNR of GW231123.

Another key implication of lensing is that it changes the inferred properties of the source. 
Under the unlensed hypothesis, GW231123 appears as an exceptionally massive and rapidly spinning binary. Once lensing is included, the inferred source-frame masses shift to lower values and the spins become less extreme. For example, with NRSur, the median component masses change from $(127.7,107.1)\mathrm{M}_\odot$ under U to $(112.7,59.5)\mathrm{M}_\odot$ under PM and $(99.2,51.8)\mathrm{M}_\odot$ under EB-I. More interestingly, the agreement among different waveform models also changes. Under the unlensed hypothesis, NRSur and EOB give broadly consistent mass posteriors, while Phenom shows a noticeable offset. Introducing an isolated point-mass lens does not remove this inconsistency; instead, EOB becomes the most discrepant of the three waveform models. However, under embedded lensing, the differences among NRSur, Phenom, and EOB are reduced to varying degrees. 
One possible explanation is that the inconsistency among waveform models under the unlensed hypothesis reflects unmodeled lensing signatures in the data. Without accounting for lensing, the waveform templates may absorb these effects by shifting the inferred source parameters toward larger masses and higher spins. This pushes the inference into a high-spin region where waveform modeling is more challenging, leading different waveform models to give different results. Once embedded lensing is included, less of the lensing effect needs to be absorbed by the intrinsic source parameters, and the discrepancies among the waveform models are consequently reduced.

A final important issue is whether such lensing evidence can also arise from an unlensed signal because of realistic detector noise and waveform systematics. To address this, we constructed an event-conditioned background by injecting a GW231123-like numerical-relativity waveform into selected real O4a H1/L1 noise. For the 100 NRSur trials, the probability for a background event to exceed GW231123 is $2$--$3\%$ within the 68\% interval obtained by propagating the Bayes-factor uncertainties. The available Phenom trials give a larger FAP, $6.5$--$8\%$, even though Phenom yields a substantially larger Bayes factor for GW231123. Moreover, all Phenom backgrounds give larger Bayes factors than their corresponding NRSur recoveries. These results show that waveform systematics can enhance not only the lensing evidence of the observed event, but also the false-positive background. 
This interpretation is consistent with recent injection studies showing that waveform-model systematics can be partially absorbed by lensing degrees of freedom, thereby producing spurious lensing support~\citep{hegde2026interplaywaveformsystematicslensing, chan2025discoveringgravitationalwaveformdistortions}.
Therefore, the absolute Bayes factor alone is not sufficient to assess the significance of a lensing candidate; it must be calibrated against realistic backgrounds using the same recovery waveform.
Nevertheless, even after accounting for waveform systematics and realistic non-Gaussian detector noise, the NRSur analysis still gives an intriguingly low FAP, leaving GW231123 as a particularly interesting lensing candidate.
In addition, we emphasize that this is an event conditioned background rather than a population averaged FAP. The injected signal is deliberately chosen to resemble GW231123 so that the background captures the susceptibility of such an extreme event to waveform systematics and realistic detector noise. 
A population drawn background would answer a different statistical question and could yield a different FAP~\citep{Goyal2025}. Because highly spinning and strongly precessing systems may be more susceptible to spurious lensing support, the event conditioned FAP reported here may overestimate the population averaged FAP.

In summary, our results show that embedded Type-I lensing provides a viable explanation for GW231123. EB-I gives the largest evidence across all three waveform models, and the NRSur event-conditioned calibration yields a low FAP. Embedded lensing also makes the inferred source properties less extreme. Taken together, these results support GW231123 as an intriguing lensing candidate.

\section*{Methods}
When the lens mass ($M_L$) and the GW frequency ($f$) meet the following condition:
\begin{equation}
\label{eq:geo_wo}
M_L \lesssim 10^{5} \mathrm{M}_{\odot}\left(\frac{f}{\mathrm{Hz}}\right)^{-1} \, ,
\end{equation}
we must use wave optics to study the lensing effect. For sources in the LVK frequency band (10 Hz to 1000 Hz), this means that wave optics effects are important for lens masses smaller than about $10^4 \mathrm{M}_\odot$.

In this case, the wave-optical effect can be evaluated by using the diffraction integral~\citep{schneider1992gravitational,10.1143/PTPS.133.137,Takahashi:2003ix}:
\begin{equation}
\label{eq:DiffInter}
F(\omega)=\frac{2 G \langle M_L \rangle \left(1+z_L\right) \omega}{\pi c^{3} i} \int_{-\infty}^{\infty} d^{2} x^\prime \exp \left[i \omega t(\boldsymbol{x^\prime})\right] , 
\end{equation}
where $F(\omega)$ is the amplification factor and $\omega$ is the angular frequency of the GW. For a binary lens inside a strong lensing galaxy, the time-delay function is:
\begin{equation}
\label{eq:TimeDelay}
\begin{split}
t(\boldsymbol{x^\prime}) = 
\frac{k}{2}\left[(1-\kappa+\gamma) x_{1}^{\prime 2} + (1-\kappa-\gamma) x_{2}^{\prime 2}\right] 
\\
- \frac{k}{2}\sum_{i=1}^{2} 
\frac{M_{L,i}}{\langle M_L \rangle}
\ln \left|\boldsymbol{x}^{\prime}-\boldsymbol{x}^{\prime i}\right|^{2} .
\end{split}
\end{equation}
Here, $\kappa$ and $\gamma$ represent the external convergence and shear provided by the strong-lensing galaxy. 
The term $\langle M_L \rangle$ is the mean mass of the two lens components in the binary-lens parameterization, and $k$ is a constant defined as $k=4 G \langle M_L \rangle(1+z_L)/c^3$, which is on the order of $10^{-5}~\mathrm{s}$ for $M_L = 1~\mathrm{M}_\odot$.
Finally, $\boldsymbol{x'}_i$ represents the position of the $i$-th lens component in units of the Einstein radius $\theta_E$ corresponding to $\langle M_L \rangle$, and $M_{L,i}$ is its mass.
Because the convergence is subject to the mass-sheet degeneracy, it cannot be independently constrained by the GW lensing signal alone. We therefore adopt $\kappa=\gamma$ as a simplifying macro-lens assumption, motivated by an isothermal model for the strong-lensing galaxy~\citep{1994A&A...284..285K}.

In this work, we consider both Type-I minimum images and Type-II saddle images. For Type I, $1-\kappa-\gamma>0$ and $1-\kappa+\gamma>0$. For Type II, these two eigenvalues have opposite signs. Under $\kappa=\gamma$, the critical value is $\gamma=0.5$. 
The EPM and EB models are used here as low-dimensional effective descriptions of the local stellar microlensing field rather than as literal one- and two-object configurations. The EPM model captures the dominant effective wave-optics response through a single effective perturber, whereas the EB model introduces a second effective perturber to allow for additional local structure.

Unlike an isolated point mass lens, we cannot solve the amplification factor $F(\omega)$ for the embedded lens analytically. 
We need to use numerical integration, which is computationally prohibitive for standard analysis methods like Markov Chain Monte Carlo or nested sampling.
To address the computational cost of the diffraction integral, we trained separate convolutional neural-network surrogates for the Type-I and Type-II embedded geometries. Their training amplification factors were generated with the corresponding numerical wave-optics solvers, following the algorithms introduced in~\citet{Shan:2022xfx, Shan:2023ngi, Shan:2024min}.
Here, it is worth noting that we trained surrogate models only for EB-I and EB-II, since EPM-I and EPM-II are recovered as the single-lens limits of their corresponding EB models.

To facilitate the learning of wave-optical signatures, we do not directly train the network on the raw diffraction integral results. Instead, we train it to learn the residual between the diffraction integral and the geometric approximation:
\begin{equation}
\label{eq:residual}
    F_\mathrm{residual} (f) = F_\mathrm{wave} - F_\mathrm{Geo} \, ,
\end{equation}
where the geometric amplification factor, $F_\mathrm{Geo}$, is defined as:
\begin{equation}
    F_\mathrm{Geo} = \sum_{i=1}^{N} \sqrt{|\mu_i|} \exp(i\pi(2 f t_i - n_i)) \, .
\end{equation}
Here, $\mu_i$, $t_i$, and $n_i$ are the magnification, time delay, and discrete phase shift of the $i$-th microlensed image, respectively. The phase shift $n_i$ takes values of 0, 1/2, or 1 for Type I (minimum), Type II (saddle), and Type III (maximum) images.

Extended Data Figure~\ref{fig:transformer} shows the neural-network structure shared by the Type-I and Type-II surrogates, which are trained independently. Their seven-dimensional conditioning vector contains the external shear, the redshifted mean lens mass, the lens mass ratio, and two lens-plane coordinates for each effective perturber; the training and test parameter ranges are summarized in Extended Data Table~\ref{tab:train}.
The main branch receives four normalized features: frequency, the logarithm of the dimensionless frequency $w=8\pi G\langle M_L^z\rangle f/c^3$, and the real and imaginary parts of $F_\mathrm{Geo}$. A DeepSets encoder maps the geometric-optics image information to 128 features. The image table is padded to a maximum of nine images, corresponding to the $5N-1$ upper bound for $N=2$ point masses in a constant external shear~\citep{An_2006}. Each image is described by its dimensionless time delay, transformed magnification $\log[1+\sqrt{|\mu|}]$, two image-plane coordinates $(x_1,x_2)$, and two indicators specifying whether it is a minimum or saddle. A fully connected network maps the seven lensing parameters to 64 features, which are combined with the image features to modulate six one-dimensional convolutional residual blocks with 64 channels, kernel size 5, and dilation factors $(1,2,16,32,64,128)$.

The network predicts the real and imaginary parts of the correction to the geometric-optics amplification factor, $F_\mathrm{residual}(f)$, from which the full result is reconstructed as $F_\mathrm{wave}(f)=F_\mathrm{Geo}(f)+F_\mathrm{residual}(f)$. The network is trained to reproduce the complex amplification factor, with an additional detector-weighted loss that emphasizes errors most relevant to the gravitational-wave signal. For each strong lensing geometry, we select the model with the best validation-set mismatch performance.

% \codex{We evaluate the trained networks on independent test samples. Extended Data Figure~\ref{fig:app2} shows the mismatches of the Type-I geometric-optics approximation and the Type-I and Type-II neural predictions relative to the corresponding full wave-optics calculations for a GW231123-like signal.}

Extended Data Figure~\ref{fig:app2} compares the mismatches of the EB-I geometric-optics approximation (blue), the EB-I neural surrogate (orange), and the EB-II neural surrogate (green) relative to the corresponding full wave-optics calculations for a GW231123-like signal. The vertical line marks the conservative single-parameter mismatch criterion $1.35/\rho^2=0.0028$ at $\rho=22$~\citep{2010PhRvD..82b4014M,2013PhRvD..87b4035B,thompson2025useinterpretationsignalmodelindistinguishability,LIGOScientific:2025rsn}. The test-set parameter space is the same as that of the training set, as shown in Extended Data Table~\ref{tab:train}.
The fractions above this threshold are $12.69\%$ for geometric optics, $0.70\%$ for the EB-I surrogate, and $0.33\%$ for the EB-II surrogate. Thus, the Type-I geometric-optics approximation is not sufficiently accurate over the parameter space considered here, while both neural networks successfully capture the wave-optics structure required for the subsequent inference.

For parameter estimation, we use the same source data, calibration, and data-analysis priors as the LVK analysis~\citep{LIGOScientific:2025rsn}. The isolated point-mass model uses a log-uniform prior on the detector-frame redshifted lens mass, $M_L^z\in[1,10^4]\,\mathrm{M}_\odot$, and $p(y)\propto y$ for $y\in[0.1,10]$. For the embedded models, $M_L^z=M_{L,1}^z+M_{L,2}^z$ denotes the detector-frame redshifted total lens mass, with $M_{L,1}^z=M_L^z/(1+q_L)$ and $M_{L,2}^z=q_LM_{L,1}^z$, and the local source position is fixed at the macro-image origin. The EPM model is represented by the one-effective-perturber limit within the surrogate domain. For the EPM approximation only, the secondary perturber is fixed to the minimum mass ratio, $q_L=0.01$, the maximum separation, $r_2=10\,\theta_E$, and $\theta_{L,2}=0$, while $M_L^z$, $\gamma$, $r_1$, and $\theta_{L,1}$ follow the corresponding embedded-model priors. 
We test the accuracy of this approximate EPM model by comparing it with the exact results obtained from numerical integration. The comparison is shown in Extended Data Figure~\ref{fig:epm_mismatch_histogram_typei_typeii}. Only 3.5\% and 4\% of the EPM-I and EPM-II samples, respectively, exceed this threshold, indicating that the errors introduced by the single-effective-perturber approximation are subdominant to the intrinsic GW waveform-model systematics~\citep{LIGOScientific:2025rsn}.
For the EB models, all binary-lens parameters are sampled. We use a log-uniform prior on $M_L^z\in[1,10^4]\,\mathrm{M}_\odot$, a log-uniform prior on $q_L\in[0.01,1]$, $p(r_i)\propto r_i$ for $r_i\in[0.1,10]\,\theta_E$, with $\theta_{L,1}\in[0,\pi/2]$ and $\theta_{L,2}\in[0,2\pi]$. The shear prior is $0\leq\gamma<0.48$ for Type I and $0.52\leq\gamma<1$ for Type II. All hypotheses are analyzed with \texttt{Bilby} using the \texttt{dynesty} sampler, and Bayes factors are computed from the corresponding evidences.

For background calibration, we use the unlensed SXS:BBH:2324 numerical-relativity waveform~\citep{Varma:2020bon}. We selected this simulation from the Simulating eXtreme Spacetimes (SXS) catalog by searching for simulations consistent with the intrinsic parameters of GW231123 at 10 Hz. It is the nearest SXS simulation to the maximum-likelihood intrinsic parameters of GW231123 and also lies within its joint intrinsic posterior support. We include multipoles $\ell=2$--$8$ and inject the waveform into 100 selected real O4a H1/L1 noise realizations. The source orientation and detector geometry are fixed to the corresponding values for GW231123. For each realization, we use the corresponding H1 and L1 PSDs and apply a common amplitude scaling factor to the injected signals in both detectors so that the resulting network SNR is $21.95$, matching that of GW231123. The recoveries use 8 s segments sampled at 1024 Hz, a 20--448 Hz likelihood band, and a Tukey analysis window with a 1 s roll-off. Each injection is recovered under the U and EB-I hypotheses using the same production NRSur parameter-estimation framework as for GW231123. Because the injections use direct numerical-relativity waveforms whereas the recoveries use the NRSur waveform model, the resulting background captures injection--recovery waveform-model differences, together with the effects of realistic non-Gaussian and non-stationary detector noise. For the auxiliary Phenom cross-check, we analyze the 18 completed paired trials from the same 100-trial background set, limiting the additional computational resources devoted to this comparison. Extended Data Figure~\ref{fig:appendix_figure3_background_injection_nrsur_best_fit} shows one representative background injection together with the corresponding NRSur and Phenom waveforms. The legend reports their mismatches relative to the injected numerical-relativity waveform after optimizing only the source-orientation angles, a common network time shift and distance, while keeping the intrinsic parameters and sky geometry fixed.

For each recovery waveform, we construct the complementary cumulative distribution function (CCDF) of $\log_{10}B^{\rm EB-I}_{\rm U}$ for the background trials and compare it with the value measured for GW231123. The FAP is obtained from the fraction of background trials with lensing evidence at least as large as that measured for GW231123. We include the uncertainty in the Bayes-factor estimates when calculating the FAP, which gives the quoted 68\% interval.

% Nature Astronomy requires separate Data availability and Code availability
% statements after Methods. They are intentionally left as comments here so
% that no manuscript content is added or rewritten.
% \section*{Data availability}
% \section*{Code availability}

\clearpage
\section*{Extended Data}

\setcounter{figure}{0}
\renewcommand{\figurename}{Extended Data Figure}

\begin{figure*}
\centering
\includegraphics[width=\textwidth]{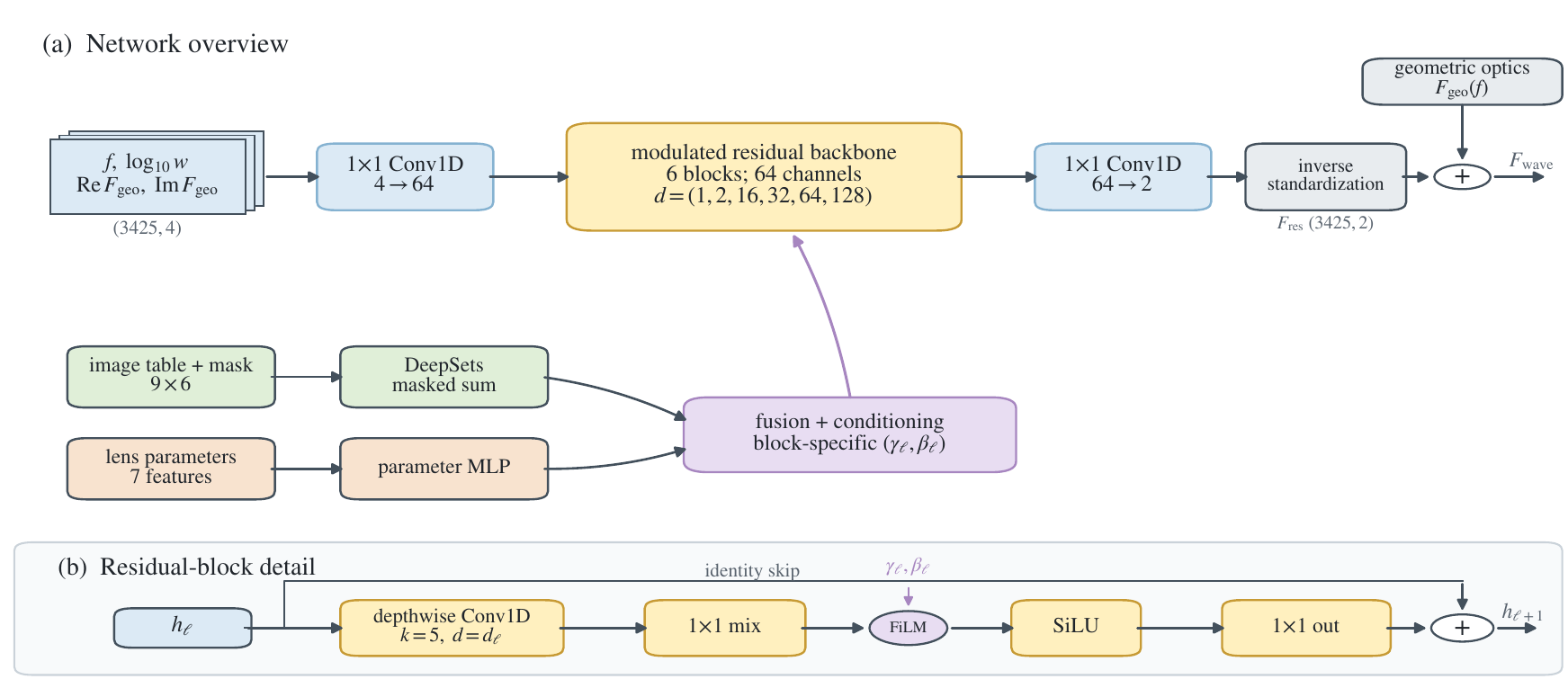}
\caption{The neural-network architecture used to predict the wave-optics amplification factor. (a) The network combines frequency-dependent geometric-optics information with the image properties and seven lensing parameters to predict the wave-optics correction $F_\mathrm{residual}(f)$. The full amplification factor is then reconstructed as $F_\mathrm{wave}(f)=F_\mathrm{Geo}(f)+F_\mathrm{residual}(f)$. (b) Structure of each residual block, which contains dilated convolution, channel mixing, feature modulation, and a skip connection. The Type-I and Type-II surrogates use the same architecture but are trained independently.}
\label{fig:transformer}
\end{figure*}

\begin{figure}
\centering
\includegraphics[width=\linewidth]{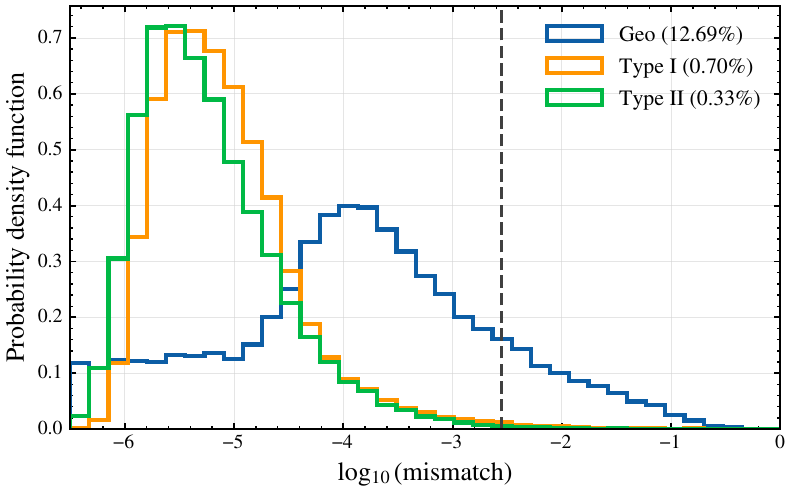}
\caption{Mismatch distributions for the Type-I geometric-optics approximation (Geo) and the Type-I and Type-II neural surrogates on independent test samples. Each prediction is compared with the corresponding full wave-optics calculation, with the mismatch weighted using the GW231123 waveform and noise spectra. The black dashed line marks the mismatch criterion $1.35/\rho^2=0.0028$ for $\rho=22$, and the legend gives the fraction of samples above this threshold.
}
\label{fig:app2}
\end{figure}

\begin{figure}
\centering
\includegraphics[width=\linewidth]{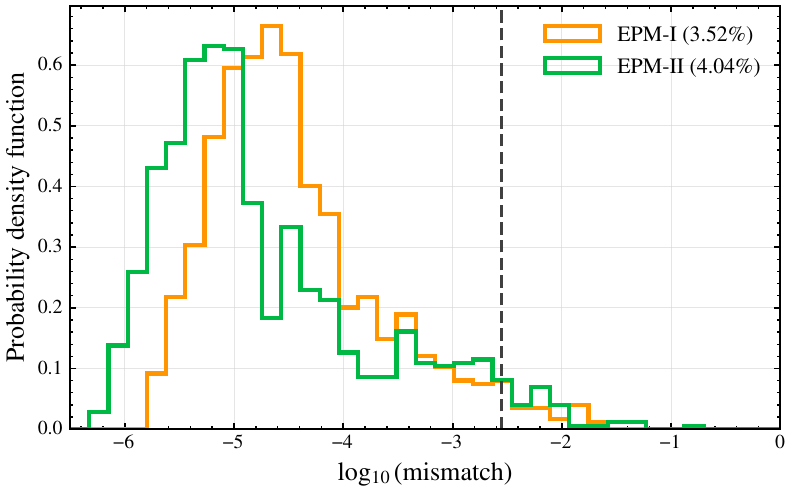}
\caption{Mismatch distributions for the approximate EPM-I and EPM-II models relative to the corresponding exact numerical wave-optics calculations. The black dashed line marks the adopted mismatch threshold $1.35/\rho^2=0.0028$ for $\rho=22$. Only 3.5\% and 4\% of the EPM-I and EPM-II samples, respectively, exceed this threshold, indicating that the errors introduced by the single-effective-perturber approximation are subdominant to the intrinsic GW waveform-model systematics~\citep{LIGOScientific:2025rsn}.}
\label{fig:epm_mismatch_histogram_typei_typeii}
\end{figure}

\begin{figure*}
\centering
\includegraphics[width=\textwidth]{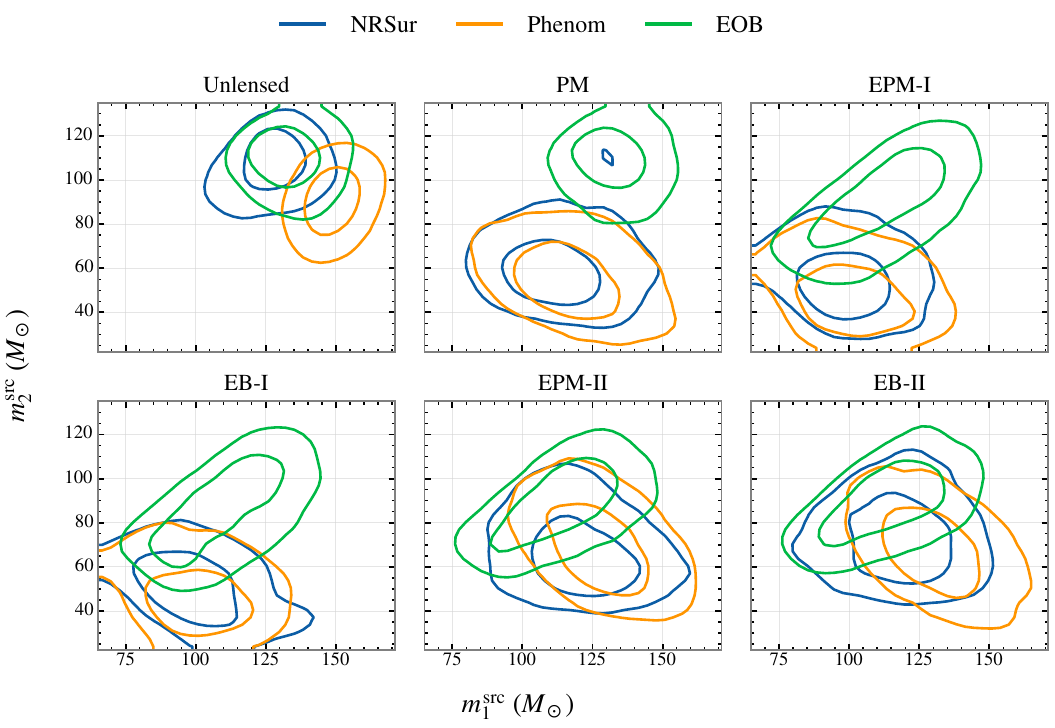}
\caption{Source-frame component-mass posteriors for U, PM, EPM-I, EB-I, EPM-II, and EB-II, ordered from left to right and top to bottom. Each panel compares NRSur, Phenom, and EOB. The contours enclose 50\% and 90\% credible regions.}
\label{fig:source_mass_appendix}
\end{figure*}

\begin{figure*}
\centering
\includegraphics[width=\textwidth]{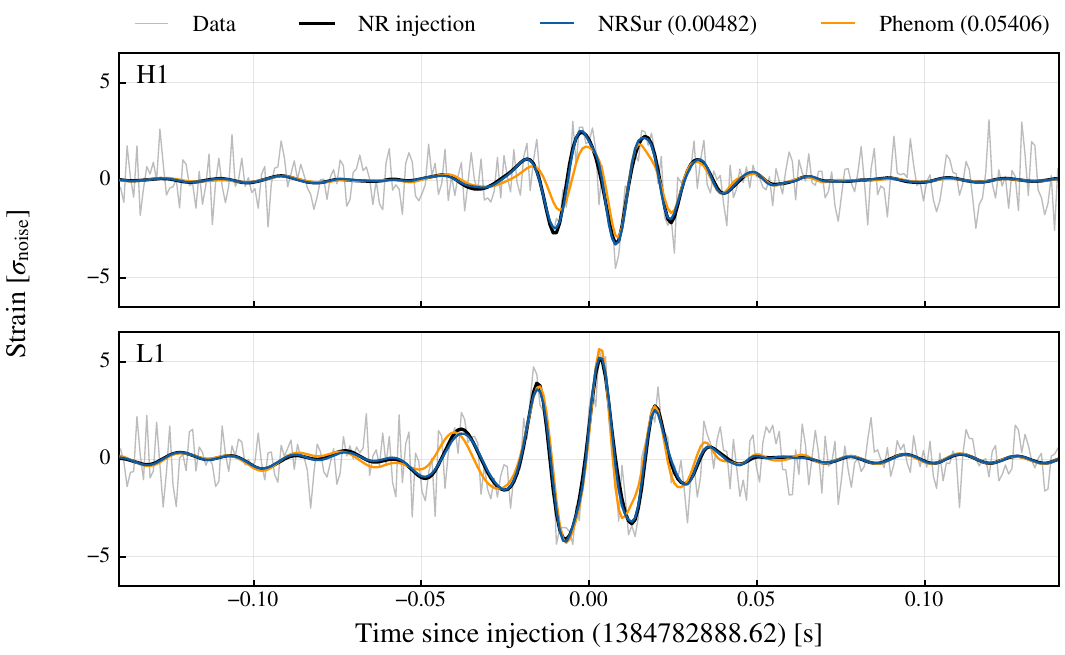}
\caption{One representative background injection together with the corresponding NRSur and Phenom waveforms. The legend reports their mismatches relative to the injected numerical-relativity waveform after optimizing only the source-orientation angles, a common network time shift and distance, while keeping the intrinsic parameters and sky geometry fixed.}
\label{fig:appendix_figure3_background_injection_nrsur_best_fit}
\end{figure*}

\setcounter{table}{0}
\renewcommand{\tablename}{Extended Data Table}

\begin{table*}[h!]
\centering
\renewcommand{\arraystretch}{1.15}
\caption{NRSur source properties under U, PM, EPM-I, and EB-I. Here $m_1^{\mathrm{src}}$ and $m_2^{\mathrm{src}}$ are the source-frame component masses, $a_1$ and $a_2$ are the dimensionless spin magnitudes of the two black holes, $\chi_p$ is the effective precession spin parameter, and $\chi_{\mathrm{eff}}$ is the effective spin component along the orbital angular momentum. Values are medians with 90\% credible intervals.}
\begin{tabular}{lcccc}
\toprule
Parameter & U & PM & EPM-I & EB-I \\
\midrule
$m_1^{\mathrm{src}}$ [$\mathrm{M}_\odot$] & $127.7^{+16.0}_{-17.1}$ & $112.7^{+25.4}_{-21.4}$ & $98.7^{+26.0}_{-22.1}$ & $99.2^{+30.9}_{-23.1}$ \\
$m_2^{\mathrm{src}}$ [$\mathrm{M}_\odot$] & $107.1^{+15.7}_{-19.4}$ & $59.5^{+30.3}_{-16.2}$ & $54.8^{+26.1}_{-18.1}$ & $51.8^{+23.0}_{-20.7}$ \\
$a_1$ & $0.90^{+0.08}_{-0.29}$ & $0.82^{+0.15}_{-0.40}$ & $0.81^{+0.16}_{-0.58}$ & $0.74^{+0.22}_{-0.50}$ \\
$a_2$ & $0.90^{+0.08}_{-0.33}$ & $0.48^{+0.46}_{-0.43}$ & $0.44^{+0.49}_{-0.40}$ & $0.54^{+0.40}_{-0.47}$ \\
$\chi_p$ & $0.76^{+0.17}_{-0.20}$ & $0.50^{+0.33}_{-0.33}$ & $0.52^{+0.33}_{-0.34}$ & $0.46^{+0.37}_{-0.30}$ \\
$\chi_{\mathrm{eff}}$ & $0.26^{+0.24}_{-0.37}$ & $0.40^{+0.21}_{-0.27}$ & $0.36^{+0.21}_{-0.38}$ & $0.41^{+0.21}_{-0.42}$ \\
\bottomrule
\end{tabular}
\label{tab:source_params_comparison}
\end{table*}

\begin{table*}[h!]
\centering
\renewcommand{\arraystretch}{1.15}
\caption{EB-I lens parameters obtained with NRSur, Phenom, and EOB. Here $\gamma$ is the external shear, $\mu$ is the macro magnification under $\gamma=\kappa$, $M^z_{L,1}$ and $M^z_{L,2}$ are the redshifted effective perturber masses, and $(x_{11},x_{12})$ and $(x_{21},x_{22})$ are their lens-plane coordinates in units of $\theta_\mathrm{E}$ defined by the mean lens mass $\langle M_L\rangle$. Values are medians with 90\% credible intervals.}
\begin{tabular}{lccc}
\toprule
Parameter & NRSur & Phenom & EOB \\
\midrule
$\gamma$ & $0.40^{+0.06}_{-0.22}$ & $0.41^{+0.06}_{-0.18}$ & $0.36^{+0.11}_{-0.30}$ \\
$\mu$ & $4.97^{+6.09}_{-3.43}$ & $5.48^{+9.31}_{-3.63}$ & $3.69^{+12.81}_{-2.55}$ \\
$M^z_{L,1}$ [$\mathrm{M}_\odot$] & $622.2^{+455.1}_{-357.2}$ & $379.6^{+700.3}_{-134.7}$ & $149.4^{+3044.6}_{-147.8}$ \\
$M^z_{L,2}$ [$\mathrm{M}_\odot$] & $52.1^{+313.9}_{-46.5}$ & $25.6^{+249.8}_{-21.3}$ & $11.1^{+533.4}_{-11.0}$ \\
$x_{11} [\theta_\mathrm{E}]$ & $1.12^{+0.69}_{-0.84}$ & $0.72^{+1.29}_{-0.53}$ & $3.01^{+5.40}_{-2.73}$ \\
$x_{12} [\theta_\mathrm{E}]$ & $2.39^{+3.04}_{-1.11}$ & $3.47^{+3.26}_{-1.97}$ & $3.88^{+4.92}_{-3.32}$ \\
$x_{21} [\theta_\mathrm{E}]$ & $-0.13^{+8.33}_{-8.06}$ & $-0.45^{+8.49}_{-7.87}$ & $-0.07^{+7.94}_{-8.01}$ \\
$x_{22} [\theta_\mathrm{E}]$ & $0.72^{+7.63}_{-9.01}$ & $1.08^{+7.28}_{-9.37}$ & $0.11^{+7.95}_{-8.18}$ \\
\bottomrule
\end{tabular}
\label{tab:lens_params}
\end{table*}

\begin{table}[h!]
    \centering
    \renewcommand{\arraystretch}{1.15}
    \caption{Training and test-set parameter ranges for the Type-I and Type-II neural surrogates. Here $\gamma$ is the external shear, $\langle M_L^z\rangle=(1+z_L)\langle M_L\rangle$ is the redshifted mean lens mass, $q_L$ is the lens mass ratio $q_L=M_{L,2}/M_{L,1}\leq1$, and $(x_{11},x_{12})$ and $(x_{21},x_{22})$ are the lens-plane coordinates of the two effective perturbers in units of the Einstein radius $\theta_{\mathrm{E}}$ defined by the mean lens mass $\langle M_L\rangle$.}
    \begin{tabular}{lcc}
    \toprule
    Parameter & Type I & Type II \\
    \midrule
    $\gamma$ & $[0,0.48)$ & $[0.52,1)$ \\
    $\langle M_L^z\rangle$ [$\mathrm{M}_\odot$] & $[0.5,5000]$ & $[0.5,5000]$ \\
    $q_L$ & $[0.01,1]$ & $[0.01,1]$ \\
    $(x_{11},x_{12})$ [$\theta_\mathrm{E}$] & $[0,10]^2$ & $[0,10]^2$ \\
    $(x_{21},x_{22})$ [$\theta_\mathrm{E}$] & $[-10,10]^2$ & $[-10,10]^2$ \\
    \bottomrule
    \end{tabular}
    \label{tab:train}
\end{table}

\begin{acknowledgments}
\noindent\emph{\textbf{Acknowledgments.—}}
    We thank Miguel Zumalacárregui and Ania Liu for their helpful comments and suggestions.
    ~H.Y. is supported by the National Natural Science Foundation of China (Grant 12573048).
    X.S. and S.M. acknowledge support from the National Science Foundation of China (Grant No. 12133005).
    X.S. acknowledges support from Shuimu Tsinghua Scholar Program (No. 2024SM199), the China Postdoctoral Science Foundation (Certificate Number: 2025M773189), the ERC grants GWSky (101167314) and GLOW (101230608).
    This material is based upon work supported by NSF's LIGO Laboratory which is a major facility fully funded by the National Science Foundation.
\end{acknowledgments}

% Nature Astronomy also requires author contributions, competing interests,
% and additional information/correspondence statements at the end. These are
% not inserted because they were not present in the supplied manuscript.

% For final Nature Astronomy submission, numerical references should be
% embedded in the manuscript file (copy the .bbl contents here). BibTeX is
% retained in this working version so the scientific content is unchanged.
\bibliography{prl_ref}

\end{document}